\documentclass[aps,prb,twocolumn,showpacs,superscriptaddress]{revtex4}
\usepackage{graphicx}
\usepackage{bm}

\usepackage{graphicx}
\usepackage{bm}

\newcommand{\beq}{\begin{equation}}
\newcommand{\eeq}{\end{equation}}

\begin{document}

\title{
Interplay of single-particle and collective degrees of freedom near
the quantum critical point}
\author{V.~A.~Khodel}
\affiliation{Russian Research Centre Kurchatov
Institute, Moscow, 123182, Russia}
\affiliation{McDonnell Center for the Space Sciences \&
Department of Physics, Washington University,
St.~Louis, MO 63130, USA}
\author{J.~W.~Clark}

\affiliation{McDonnell Center for the Space Sciences \&
Department of Physics, Washington University,
St.~Louis, MO 63130, USA}
\author{M.~V.~Zverev}
\affiliation{Russian Research Centre Kurchatov
Institute, Moscow, 123182, Russia}
\date{\today}
\begin{abstract}
Competing scenarios for quantum critical points (QCPs) of strongly
interacting Fermi systems signaled by a divergent density of
states at zero temperature are contrasted.  The conventional
scenario, which enlists critical fluctuations of a collective
mode and attributes the divergence to a coincident vanishing of
the quasiparticle strength $z$, is shown to be incompatible
with identities arising from conservation laws prevailing in the
fermionic medium.  An alternative scenario, in which the topology
of the Fermi surface is altered at the QCP, is found to explain
the non-Fermi-liquid thermodynamic behavior observed experimentally in
Yb-based compounds close to the QCP.  It is suggested that combination
of the topological scenario with the theory of quantum phase
transitions will provide a proper foundation for analysis of
the extended QCP region.
\end{abstract}

\maketitle

\paragraph{Introduction.}
Fundamental understanding of the behavior of Fermi systems in
the vicinity of quantum phase transitions persists as one
of the most challenging objectives of condensed-matter research.
As it involves second-order transitions occurring at a critical
density $\rho_c$, the problem is even more difficult than in the
classical regime, since the description of quantum fluctuations
entails a new critical index, the dynamical critical
exponent.\cite{sachdev,loh} In several heavy-fermion
metals---notably Yb-based compounds\cite{gegenwart}---critical
temperatures $T_N(B)$ can be driven to zero by extremely weak
magnetic fields $B$, creating a quantum critical
point (QCP).  It is commonly believed that low-temperature
fluctuation contributions to the free energy, specific heat
$C(T)$, and other thermodynamic quantities must then follow
power laws in $T$, the Sommerfeld ratio $C(T)/T$ being divergent
at $T\to 0$.

From a pedestrian standpoint, such non-Fermi-liquid (NFL) behavior
must extend some distance from the QCP, implying separation at
$T=0$ of a domain of magnetic ordering from a Fermi-liquid
(FL) regime---as is indeed the case in the heavy-fermion
metal YbAgGe.\cite{bud'ko}  However, this example is unique;
otherwise, the two phases seem to abut each other at the
quantum critical point (QCP).  Since the standard FL formalism is
applicable on the FL side of the QCP, $C(T)/T$
is proportional to the effective mass $M^*$ in this region
and it follows that $M^*$ must diverge at the QCP.

Conventional arguments that quasiparticles in Fermi liquids
``get heavy and die''\cite{coleman} at the QCP commonly
employ the textbook formula
\beq
{M\over
M^*}=z\left[1+{1\over v^0_F}\left({\partial\Sigma(p,\varepsilon)
\over\partial p}\right)_0\right],
\label{meffz}
\eeq
where $v^0_F=p_F/M$ and the derivative is evaluated at $p=p_F$ and
$\varepsilon=0$, single-particle (sp) energies being referred
to the chemical potential $\mu$.  The factor
$z=\left[1-\left(\partial \Sigma(p,\varepsilon)/
\partial \varepsilon\right)_0\right]^{-1}$ is the quasiparticle
weight of the sp state at the Fermi surface.  The conventional
belief, traced back to Ref.~\onlinecite{doniach}, holds that the
divergence of $M^*$ at the QCP is caused by the vanishing
of the $z$ factor in Eq.~(\ref{meffz}), stemming from the
divergence of the derivative $\left(\partial
\Sigma(p,\varepsilon;\rho_c)/ \partial \varepsilon\right)_0$
at implicated second-order phase transition points.

However, this scenario is problematic.  As will be seen, the $z$-factor
does not vanish at the points of second-order phase transitions.
It will be argued that the divergence of the density of states
$N(T)$ at the QCP is in fact associated with a rearrangement of
{\it single-particle degrees of freedom},\cite{prb2008} rather
than with critical fluctuations. Even so, the divergence
of $N(T)$ at the QCP does give rise to some second-order phase
transitions, occurring at $T=T_N$ in the vicinity of the QCP.
Accordingly, the full pattern of phenomena in the temperature
interval from 0 to $T\geq T_N$ in the QCP region is determined
by an intricate interplay between the two mechanisms.  The
present analysis is limited to the disordered side of the QCP
where the impact of sp degrees of freedom is decisive, while
the role of critical fluctuations is suppressed.  Other regions
of the phase diagram will be analyzed elsewhere.

\paragraph{Fault lines of the conventional scenario for the divergence
of $M^*$.}   We begin by exposing inconsistencies of the standard
derivation leading to divergence of $\left(\partial \Sigma(p,\varepsilon)/
\partial \varepsilon\right)_0$ in the vicinity of second-order phase
transitions.  This derivation is based on a fundamental relation
of many-body theory,
\beq
{\partial \Sigma_{\alpha\delta}(p,\varepsilon)\over \partial \varepsilon}
=-{1\over 2}\int U_{\alpha\delta\gamma\beta}({\bf p},\varepsilon,{\bf l},
\varepsilon_1) {\partial G_{\beta\gamma}(l,\varepsilon_1)\over\partial\varepsilon_1}{d{\bf l}d \varepsilon_1\over (2\pi)^4i},
\label{ward}
\eeq
where $U$ is the totality of diagrams of the scattering amplitude, irreducible in the longitudinal particle-hole channel.

The treatment in question retains only the pole part $G^q$ of the
sp Green function $G=zG^q+G^r$ and a singular part of the diagram
block $U$ that is supposedly responsible for the divergence of
$\left(\partial \Sigma(p,\varepsilon)/\partial \varepsilon\right)_0$.
By definition of the block $U$,
there is no the direct spin-fluctuation
contribution to
$\left(\partial \Sigma(p,\varepsilon)/\partial \varepsilon\right)_0$,
while the exchange term has the form
$U_{\alpha\delta\gamma\beta}({\bf p},\varepsilon,{\bf l},\varepsilon_1)
=g^2 {\bm \sigma}_{\alpha\beta}{\bm \sigma}_{\gamma\delta} \chi(q,q_0)$
involving the spin susceptibility $\chi(q,q_0)$,
where $q=|{\bf p}-{\bf l}|$, $q_0=\varepsilon-\varepsilon_1$,
and $g$ is a dimensionless effective coupling constant.

On the FL side of the QCP, $\Sigma_{\alpha\delta}(p,\varepsilon)
=\Sigma(p,\varepsilon)\delta_{\alpha\delta}$ and
${\rm Im}\,G^q(p,\varepsilon)=-2\pi{\rm sgn}[\epsilon(p)]
\delta(\varepsilon-\epsilon(p))$.
In accord with the FL perspective and conventions, we now
write
$U_{\alpha\delta\gamma\beta} \equiv U_o\delta_{\alpha\delta}
\delta_{\gamma\beta}
+U_s{\bm \sigma}_{\alpha\delta}{\bm \sigma}_{\gamma\beta}$,
and the single component $U_o$ enters Eq.~(\ref{ward}), yielding
\beq
{\partial\Sigma(p,\varepsilon)\over \partial  \varepsilon}=-
\int  U_o (p,\varepsilon,{\bf l},\varepsilon_1)
{\partial G(l,\varepsilon_1)\over\partial \varepsilon_1}{d{\bf l}
d\varepsilon_1\over (2\pi)^4i}.
\label{der0}
\eeq
Applying the identity
${\bm \sigma}_{\alpha\beta}{\bm \sigma}_{\gamma\delta}=
{3\over 2}\delta_{\alpha\delta}\delta_{\gamma\beta}-
{1\over 2}{\bm \sigma}_{\alpha\delta}{\bm \sigma}_{\gamma\beta}$,
we have $U_o(q,q_0)=(3/2)g^2\chi(q,q_0)$ and
Eq.~(\ref{ward}) reduces to\cite{mig,doniach}
\beq
\left({\partial \Sigma(p,\varepsilon)\over
\partial \varepsilon}\right)_0\sim -g^2z \left({dp\over d\epsilon(p)}\right)_0
\int \chi(q,q_0=0){qdq\over \pi^2}.
\label{ints}
\eeq

A key assumption made in Ref.~\onlinecite{doniach}, and generally
adopted in subsequent treatments, is the Ornstein-Zernike (OZ) form
\beq
\chi(q,q_0=0)\equiv \chi_{\rm OZ}(q)= {4\pi\over \xi^{-2}+q^2}
\label{rsp}
\eeq
for the static correlation function $\chi(q)$, with the correlation
length $\xi$ diverging at the critical point.
Inserting this ansatz into Eq.~(\ref{ints}) together with
$v_0=\left( d\epsilon(p)/dp\right)_0=v^0_FM/M^*$,
one arrives at
\beq
\left({\partial \Sigma(p,\varepsilon)\over \partial\varepsilon}\right)_0\sim -{g^2\over v^0_F}\ln(p_F\xi),
\label{mds}
\eeq
which implies that $\left({\partial \Sigma(p,\varepsilon;\rho)/ \partial
\varepsilon}\right)_0$ diverges at $\xi\to \infty$.  However,
the applicability of the OZ approximation to evaluation of
$\left(\partial \Sigma(p,\varepsilon)/\partial \varepsilon\right)_0$
in the QCP domain has never been proved.

This deficiency exhorts us to check the compatibility of the
OZ approximation in homogeneous matter with a set of identities
involving the derivative $\left(\partial \Sigma(p,\varepsilon)/\partial
\varepsilon\right)_0$, all having the same structure as
Eq.~(\ref{ward}).
In so doing, we observe that Eq.~(\ref{ward}), which follows from
particle-number conservation with the aid of the scalar gauge
transformation $\Psi(t)\to \Psi(t) e^{i V t}$, is but one instance of a
class of similar identities.\cite{migdal}  Any conservation law
existing in the medium generates a corresponding identity involving
$\left(\partial \Sigma(p,\varepsilon)/ \partial \varepsilon\right)_0$.
For example, momentum conservation in homogeneous, isotropic matter,
associated with the vector gauge transformation
$\Psi(t)\to \Psi(t) e^{i{\bf p}{\bf A}t}$, results in the
well-known Pitaevskii relation\cite{trio}
\beq
{\partial\Sigma(p,\varepsilon)\over \partial  \varepsilon}=-
\int  U_o (p,\varepsilon,{\bf l},\varepsilon_1)
{\partial G(l,\varepsilon_1)\over\partial \varepsilon_1}
{\left({\bf p}{\bf l}\right)\over p^2}{d{\bf l}
d\varepsilon_1\over (2\pi)^4i}.
\label{derm}
\eeq
Analogously, in the model of Ref.~\onlinecite{doniach}
the spin operator $\sigma_3$ commutes with the Hamiltonian,
and the gauge transformation $\Psi(t)\to \Psi(t) e^{i\sigma_3 V t}$
leads to the relation
\beq
{\partial\Sigma(p,\varepsilon)\over \partial  \varepsilon}=-
\int  U_s( {\bf p},\varepsilon, {\bf l},\varepsilon_1)
{\partial G(l,\varepsilon_1)\over\partial \varepsilon_1}
{d{\bf l}d\varepsilon_1\over (2\pi)^4i}.
\label{ders}
\eeq
Even more conservation laws exist in nuclear and dense quark matter,
each providing an identity like Eq.~(\ref{ward}).

The standard manipulations applied to relation (\ref{ward}),
leading to the result (\ref{mds}) via ansatz (\ref{rsp}), can now
be repeated for any such conservation identity.  Irrespective of
which identity is chosen, a divergent result is obtained for
$\left(\partial\Sigma(p,\varepsilon;\rho_c)/\partial\varepsilon\right)_0$.
Importantly, however, the signs of the divergent components
of this derivative {\it do} depend on the choice made.  For example,
in the case of critical spin fluctuations, adoption of the anzatz
(\ref{rsp}) results in divergence of the derivative
$\left(\partial\Sigma(p,\varepsilon;\rho_c)/\partial\varepsilon\right)_0$
whether Eq.~(\ref{der0}) or Eq.~(\ref{ders}) is adopted, but
{\it different} signs are delivered, since in the OZ approximation
the blocks $U_o$ and $U_s$ have {\it opposite} signs.  Thus,
Eq.~(\ref{der0}) provides an ``acceptable'' negative sign, whereas
Eq.~(\ref{ders}) gives a ``wild'' {\it positive} sign (and a meaningless
limit for $z$).  In the case of short-wave-length fluctuations with
critical wave number $q_c$, a similar wild result is obtained
from Eq.~(\ref{derm}) because the prefactor of the divergent part of
$\left(\partial\Sigma(p,\varepsilon)/\partial\varepsilon\right)_0$
differs from that derived\cite{cgy} from Eq.~(\ref{der0}) by
a factor $\cos\theta_c=1-q^2_c/2p^2_F$.  Since the nonsingular
components of the block $U$ are incapable of compensating the
divergent OZ contributions to
$\left(\partial\Sigma(p,\varepsilon;\rho_c)/\partial\varepsilon\right)_0$,
we must conclude that the result (\ref{mds}) is fallacious, and
that more sophisticated methods must be applied to clarify
the situation in the critical-point region.

We call attention here to the situation for classical second-order
phase transitions, where the OZ correlation function (\ref{rsp})
is altered by scattering of the fluctuations themselves.\cite{migjun}
As a result, the actual correlation function $\chi(r,\rho_c)$
decays more rapidly at large distance $r$ than
$\chi_{\rm OZ}(r,\rho_c) \propto 1/r$.  In momentum space,
$\chi(q \to 0,q_c)$ behaves\cite{migjun} as $1/q^{2-\eta}$, with
$\eta >0$, compared with $\chi_{\rm OZ}(q) \propto 1/q^2$.
If a similar alteration of $\chi(q\to 0,\rho_c)$ occurs
at $T \to 0$, then the integration leading to (\ref{mds}) is saturated
at $\rho \to \rho_c$, ensuring that $z(\rho_c) \neq 0$.

\paragraph{Topological scenario for the QCP.}
With the condition $z(\rho_c)=0$ ruled out, the effective mass in
Eq.~(\ref{meffz}) can only diverge at a density $\rho_\infty$
where the factor in square brackets, or equivalently the group velocity,
changes sign.  Such a QCP can be examined based on the FL
equation\cite{trio}
\beq
{\bf v}({\bf p}) = {\partial\epsilon({\bf p})\over \partial {\bf p}}=
{\partial\epsilon^0_{\bf p}\over \partial {\bf  p}}
+\int f({\bf p},{\bf p}_1)
{\partial n({\bf p}_1)\over \partial {\bf p}_1}
d\tau_1
\label{lansp}
\eeq
where $\epsilon^0_{\bf p}$ is the bare sp spectrum
and $d\tau$ is the volume element in 3D or 2D momentum space.
The $T=0$ group velocity, being a continuous function of the interaction
function $f({\bf p},{\bf p}_1)$, changes its sign on the Fermi
surface at the critical density $\rho_{\infty}$.  In 3D homogeneous
matter the critical condition is
\beq
1=f_1(p_{\infty},p_{\infty})p_{\infty}M/3\pi^2,
\label{qcp}
\eeq
where $f_1$ is the first harmonic of $f$ and $p_\infty =
(3\pi^2 \rho_\infty)^{1/3}$.  In this scenario, the QCP is
associated with a rearrangement of {\it single-particle degrees
of freedom}; no collective parameter is involved, and the symmetry
of the ground state is not broken.  Such {\it topological} phase
transitions, induced by the interactions between quasiparticles,
have been discussed for over two decades.\cite{prb2008}

For a homogeneous medium there are in general two ways to realize
a divergent density of states $N(0,\rho_\infty)$.  Both options
are associated with bifurcation points $p_b$ of the equation
$\epsilon(p,\rho_{\infty})=0$.  As a condition for the divergence
of the effective mass $M^*$, Eq.~(\ref{qcp}) refers to the case\cite{ckz}
$p_b = p_F$ in which the sp spectrum $\epsilon(p)$ has an inflection
point.  In the second option, where $p_b \neq p_F$, $M^*$ remains finite,
while $N( 0,\rho_{\infty})$ diverges due to vanishing of the group
velocity at the bifurcation point.

Thus far we have dealt only with homogeneous systems. A
comparable analysis of NFL behavior of heavy-fermion metals
must include the effects of anisotropy, which are of special
importance in the QCP region.  An early study of
topological phase transitions in anisotropic electron systems
in solids, induced by electron-electron interactions,
was carried out in Ref.~\onlinecite{solid}.

Here it will be instructive to address the 2D electron liquid
in a quadratic lattice, assuming the QCP electron Fermi line
to be approximately a circle of radius $p_{\infty}$, with the
origin shifted to $(\pi/a,\pi/a)$. Since the group velocity
$v_n({\bf p})=\partial\epsilon({\bf p})/\partial p_n$ now
has a well pronounced angular dependence, the topological
{\it anisotropic} QCP is to be specified by the vanishing of
$v_n(p,\phi,T=0,\rho_{\infty})$ at the single point $p=p_{\infty}, \phi=0$.
On the FL side of the QCP where $\partial v_n(p,\phi)/\partial\phi>0$,
one has
\beq
v_n(p,\phi;T=0,\rho)
=v_n(p,\phi=0)+a_{\phi}\phi^2+a_{\rho} (\rho-\rho_{\infty}),
\label{vn}
\eeq
with $v_n(p,\phi=0)=a_p(p-p_{\infty})^2 $ as in the inflection-point
case treated in Ref.~\onlinecite{ckz}.  The QCP density of states
\beq
N(T,\rho) \propto
\int n(\epsilon)(1-n(\epsilon))
{d\epsilon\,d\phi\over v_n(p(\epsilon),\phi;T,\rho)},
\label{dst}
\eeq
which determines the specific heat $C(T)=TdS/dT\sim TN(T)$ and
thermal expansion coefficient $\beta(T)\sim -\partial
S(T,\rho)/\partial P\sim -T\partial N(T,\rho)/\partial\rho$, is
evaluated utilizing Eq.~(\ref{vn}). Following analytic integration
over $\phi$, we have
\beq
C(T,\rho_{\infty}) \propto \int n(\epsilon)(1-n(\epsilon)){d\epsilon
\over v_n^{1/2}(p(\epsilon),\phi=0;T)}\propto T^{2/3},
\label{ctq}
\eeq
\beq
\beta(T,\rho_{\infty}) \propto \int n(\epsilon)(1-n(\epsilon))
{ v_n'(\rho_\infty)d\epsilon\over v_n^{3/2}(p(\epsilon),\phi=0;T)}
\propto O(1),
\label{thq}
\eeq
where $v_n'(\rho)\equiv \partial v_n(\rho)/\partial \rho = a_\rho$
by Eq.~(\ref{vn}).  In arriving at the overall temperature behavior,
we have introduced the estimate\cite{ckz} $v_n(p,\phi=0;T)\propto T^{2/3}$
stemming from Eq.~(\ref{vn}) and valid in the relevant interval
$\epsilon\simeq T$.  The results (\ref{ctq}) and (\ref{thq}) may
be combined to determine the behavior of the Gr\"uneisen ratio,
\beq
\Gamma(T,\rho_{\infty})
=\beta(T,\rho_{\infty})/C(T,\rho_{\infty})\propto T^{-2/3},
\label{grq}
\eeq
at variance with the FL result $\Gamma \propto O(1)$.

Imposition of an external magnetic field greatly enlarges the scope
of challenging NFL behavior, as reflected in the {\it magnetic}
Gr\"uneisen ratio
$\Gamma_{\rm mag}(T,B)=-(\partial S(T,B)/\partial B)/C(T,B)$.
We now analyze this key quantity within the topological scenario,
again following the path established in Refs.~\onlinecite{ckz,shag}.
The original Fermi line is split into two, with consequent modification
of field-free relations such as (\ref{lansp}) and (\ref{dst}) through
the appearance of the sum of quasiparticle occupancies $n_\pm(\epsilon)
= \left\{ 1 + \exp\left[(\epsilon \pm \mu_e B)/T\right]\right\}^{-1}$.
As a consequence, $v_n^{-1}$ is replaced by the sum of quantities
$\left(\partial \epsilon(p,\phi)/\partial p_n\right)^{-1}$
evaluated at $\epsilon(p,\phi)\pm \mu_eB$.  Analytic integration
over $\phi$ still goes through and yields the sum of square
roots of these quantities.  In the limit $T\gg\mu_eB$, terms
in this sum linear in $r=\mu_eB/T$ cancel each other, such that
the net result is proportional to $r^2$, leading to
\beq
\Gamma_{\rm mag}(T  \gg
\mu_eB)\propto T^{-2}.
\label{grtb}
\eeq
More specifically, upon integration over $\phi$ in the field-perturbed
formulas for $C(T,B)$ and $S(T,B)$, the ensuing expressions involve
the sum $1/(\epsilon+\mu_eB)^{-1/3}+1/(\epsilon-\mu_eB)^{-1/3}$, multiplied
by a factor containing only $n(\epsilon)$. Integrating over the
dimensionless variable $y=\epsilon/T$, the ratio $S(B,T)/C(T)$
is determined as a function of $r^2$ only, yielding the result
(\ref{grtb}) at $r\ll 1$.

In the opposite limit $r \gg 1$, the density of states $N(T=0,B)$
diverges at a critical magnetic field $B_{\infty}$, where
the function $v_n(p,\phi,T=0, B_{\infty})$ vanishes on {\it one
of the two Fermi lines} $p^{\pm}(\phi)$ specified by
\beq
\epsilon(p^{\pm},\phi)\pm\mu_eB=0.
\label{fl}
\eeq
The field-induced splitting that alters the relevant
Fermi-surface group velocity can be compensated---for example,
by doping---thereby providing the means for driving the system toward
the QCP.

At $B>B_{\infty}$, the critical quantity $v_n(p,\phi=0;T=0,B)\equiv v_n(B)$
becomes positive and FL behavior is recovered, as in the isotropic
case.\cite{shag,ckz}  To evaluate the critical index specifying
the divergence of the density of states $N(T=0,B\to B_{\infty})
\propto v_n^{-1/2}(B)$, we calculate the spectrum from Eq.~(\ref{fl})
[as in Ref.~\onlinecite{ckz}] and insert the result into Eq.~(\ref{fl}),
obtaining $v_n (B)\propto (B-B_{\infty})^{2/3}$ and
$N(T=0,B)\propto (B-B_{\infty})^{-1/3}$.  Thus
\beq
C(T\to 0,B)=S(T\to 0,B)\propto T(B-B_{\infty})^{-1/3}
\label{cthq}
\eeq
and $\beta(T\to 0) \propto T(B-B_\infty)^{-1}$, so that $\Gamma(T\to 0)
\propto (B-B_\infty)^{-2/3}$.  Importantly, we arrive at
\beq
\Gamma_{\rm mag}(T\to 0)={1\over 3} (B-B_{\infty})^{-1}.
\label{grhq}
\eeq
Such a divergence was first predicted within scaling theory,\cite{zhu}
in which the peak of $\Gamma_{\rm mag}(T=0,B)$ is located at $B_c$,
the end point of the line $T_N(B)$ where $T_N(B_c)=0$.  In the
topological scenario, $B_{\infty}$ may or may not coincide with $B_c$;
this issue will be pursued in further work.

\paragraph{Discussion.}
The results (\ref{ctq})--(\ref{grtb}), (\ref{cthq}), and
(\ref{grhq}) are in agreement with available experimental
data\cite{steglich1,steglich2, tokiwa,oeschler} obtained by
the Steglich group in comprehensive studies of the thermodynamic
properties of Yb-based heavy-fermion metals.  These data also provide
a test of modern phenomenological scaling theories of the
QCP.\cite{zhu,senthil}  The outcome of this test, as aired in
Refs.~\onlinecite{tokiwa,senthil}, is that no single model
based on 2D or 3D fluctuations can describe these data, which
require the following set of critical indexes having low
probability: dimensionality $d=1$, correlation-length exponent
$\nu=2/3$, and dynamical exponent $\gamma =3/2$.

Recent studies of peaks in the specific heat $C(T,B)$ in
Yb-based compounds reveal another difficulty confronting
the phenomenological theory of second-order phase transitions
in the QCP region.  According to this theory, at $B=0$ the
difference $T-T_N$ is the single relevant parameter determining
the structure of the fluctuation peak of the Sommerfeld ratio
$C(T)/T$.  However, comparative analysis of corresponding
experimental data\cite{oeschler,steglich1} in YbRh$_2$Si$_2$
and YbRh$_2$(Si$_{0.95}$Ge$_{0.05}$)$_2$ shows clearly that
the structure of this peak {\it is not universal}.  As the
QCP is approached, the peak gradually shrinks to naught, again
bringing into question the applicability of the spin-fluctuation
scenario in its vicinity.

Thus, while the spin-fluctuation mechanism remains applicable
at {\it finite} $T\simeq T_N(B)$, it becomes {\it inadequate}
at the QCP itself. Accordingly, the relevant critical indexes
of scaling theory must be inferred anew from appropriate experimental
data on the {\it fluctuation peak} located at $T_N(B)$. Furthermore,
the existing description of thermodynamic phenomena in the
extended QCP region, including the peak at $T_N(B)$, must be
revised by integrating the topological scenario with the theory
of quantum phase transitions.\cite{sachdev}

The posited suppression of critical fluctuations in
YbRh$_2$(Si$_{0.95}$Ge$_{0.05}$)$_2$, which is situated extremely
close to the QCP in the sense that $B_c=0.027$ T, conflicts with the
conclusion of Ref.~\onlinecite{steglich3} that the system is on the
verge of a ferromagnetic instability.  This assertion is based
on extraction of the Stoner factor from measurements of the
Sommerfeld-Wilson (SW) ratio $R_{\rm SW}(T)\propto \chi(T)/C(T)$.
However, such an extraction is straightforward only in homogeneous
matter, where the magnetic part of the Hamiltonian is specified
by the Bohr magneton $\mu_B$.  In dealing with electron systems
of solids, this strategy is inconclusive unless a reliable replacement
$\mu_{\rm eff}$ for $\mu_B$ is known.  The authors of
Ref.~\onlinecite{steglich3} have chosen the effective Bohr
magneton $\mu_{\rm eff}$ to be 1.4--1.6 $\mu_B$, as determined
from data on the magnetic susceptibility itself.  Such a choice
suffers from double counting.  If instead one uses the value
$\mu_{\rm eff}=4.54\mu_B$ appropriate for the atomic state of
Yb$^{3+}$, then the Stoner factor derived from the data
remains below 3, and the conflict is resolved.

\paragraph{Conclusion.}
The conventional view of quantum critical phenomena,
in which the quasiparticle weight $z$ vanishes at points of
related $T=0$ second-order phase transitions, is incompatible with
a set of identities based on gauge transformations associated
with prevailing conservation laws.  We have traced the failure
of the standard scenario to the inapplicability of the
Ornstein-Zernike form $\chi^{-1}(q)=q^2+\xi^{-2}$ for the
static correlation function $\chi(q)$ in the limit $\xi\to \infty$.
We have discussed an alternative topological scenario and
demonstrated that its predictions for the thermodynamics of
systems on the disordered side of the QCP are in
agreement with available experimental data.  Based on these
data, we infer that close to the QCP the role of single-particle
degrees of freedom is paramount, while the effects of critical
fluctuations build up on the ordered side as the system moves
away from the QCP.

We thank A.~Alexandrov, V.~Galitski, and V.~Shaginyan for fruitful
discussions.  This research was supported by the McDonnell Center for
the Space Sciences, by Grants Nos.~NS-3004.2008.2 and 2.1.1/4540
from the Russian Ministry of Education and Science, and by Grants
Nos.~07-02-00553 and 09-02-01284 from the Russian Foundation
for Basic Research.

\end{document}